%% file: hepthversion.tex
\documentclass[12pt,a4paper]{article}
\usepackage{a4wide}
\usepackage[centertags]{amsmath}
\usepackage{amssymb}
\usepackage{epsfig}
\usepackage{psfig}
\usepackage{cite}

\begin{document}
\thispagestyle{empty}
\begin{flushright}
                     hep-th/0412066 \\
IPPP/04/82\\
DCPT/04/164
\end{flushright}
\vskip 2cm

\begin{center}
{\huge Bottom Up Meets Heterotic
  Strings$^{\mbox{\normalsize 1}}$
\footnotetext[1]{Based 
  on talks 
  given by A.\ Wingerter at String Phenomenology 2004, Ann Arbor,
  1--6 August 2004, and by S.\ F\"orste  at the 37th International
  Symposium Ahrenshoop on the Theory of Elementary Particles, Berlin,
  23--27 August 2004.}}
\vspace*{5mm} \vspace*{1cm}
\end{center}
\vspace*{5mm} \noindent
\vskip 0.5cm
\centerline{\bf Stefan F\"orste$^a$ and Ak{\i}n Wingerter$^b$}
\vskip 1cm
\centerline{$^a$ \em                                                  
Institute for Particle Physics Phenomenology (IPPP)}           
\centerline{\em South Road, Durham DH1 3LE, United Kingdom}
\vskip 0.5cm           
\centerline{$^b$  \em                                   
Physikalisches Institut der Universit\"at  Bonn}    
\centerline{\em Nussallee 12, 53115 Bonn, Germany} 
\vskip2cm

\centerline{\bf Abstract}

\noindent In this talk we argue that a certain class of heterotic orbifolds can
be the underlying fundamental theory for recently studied field theory
GUTs. In addition we demonstrate that symmetric heterotic ${\mathbb
  Z}_2 \times {\mathbb Z}_2$ orbifolds can give rise to three
generation models if quantised Wilson lines are switched on.

\newpage

\section{Introduction}

Recently it was realised by model builders that the concept of extra
dimensions leads to phenomenologically interesting models where the
standard model gauge group merges into a grand unified group only
after an extra dimension opens up. This class of models we view as
bottom up theories. We will briefly summarise the concept in section
one. 

In the remaining and main part of the talk we focus on
compactifications of heterotic E$_8 \times$E$_8$ strings on orbifolds
and argue that ${\mathbb Z}_N\times {\mathbb Z}_M$ orbifolds naturally
provide the picture obtained in the bottom up approach. For simplicity
we restrict our attention to ${\mathbb Z}_2 \times {\mathbb Z}_2$
orbifolds. In order to reduce the number of generations to three,
Wilson lines are required. We demonstrate at an example that in this way 
one can obtain three generation models.

We should emphasise that the historical order has been reversed in
this talk. Orbifolds have been studied first (and long time ago) in
the context of string theory and only recently were `rediscovered' by
model builders.
                                            
This talk is based on results reported in \cite{Forste:2004ie}.

\section{Bottom Up: Orbifold GUTs}

\begin{figure}
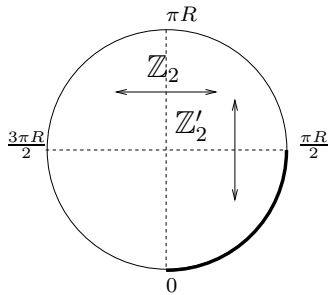

\begin{center}
\input orbicircle.pstex_t
\end{center}
\caption{The double arrows indicate the action of ${\mathbb Z}_2\times
  {\mathbb Z}_2^\prime$ on the circle. Any point on the circle can be
  mapped into the right lower quarter segment -- the fundamental
  domain is an interval of length $\pi R/2$.}
\label{fig:orbicirc}
\end{figure}

The concept of extra dimensions allows for the possibility that a
grand unified gauge group is realised only in a higher dimensional
space. Let us briefly illustrate this at a simple
example \cite{Kawamura:2000ev}. We consider the case
of one extra dimension which is compactified on an orbicircle
$S^1/{\mathbb Z}_2 \times {\mathbb Z}_2^\prime$, see figure
\ref{fig:orbicirc}.  
The orbicircle is equivalent to an interval of length $\pi R/2$. As
our five dimensional theory we take the minimal supersymmetric
extension of SU(5) gauge theory. If the extra dimension were
compactified on a circle we would get ${\cal N} = 2$ supersymmetry in
four dimensions. For the interval the situation is different. For a
point in the bulk of the interval the ${\mathbb Z}_2 \times {\mathbb
  Z}_2 ^\prime$ relates the value of a field at that point to the
value at the image points. The ends of the interval are however
invariant under elements of the orbifold action and hence we obtain a
projection there. In particular one can embed the orbifold action into
the field space such that ${\mathbb Z}_2$ projects the ${\cal N}=2$
gauge multiplet to an ${\cal N}=1$ multiplet and ${\mathbb
  Z}_2^\prime$ projects the SU(5) multiplet to an
SU(3)$\times$SU(2)$\times$U(1) multiplet. Massless fields in four
dimensions are obtained from modes which do not depend on the extra
dimension and hence we obtain ${\cal N}=1$
SU(3)$\times$SU(2)$\times$U(1) gauge theory in four dimensions. In
addition there are massive states whose mass is an integer times the
compactification scale. The standard model matter is put into the five
dimensional model either at the fixed points or into the bulk (for
more details see \cite{Kawamura:2000ev}). 

Orbifolds with one or two extra dimensions have received a lot of
attention in the recent past, e.g.\cite{Kawamura:2000ev,
  Altarelli:2001qj, Hall:2001pg, Kawamoto:2001wm, 
  Hebecker:2001wq, Asaka:2001eh}. (For a review and more references
see \cite{review}.) The reason is that these constructions allow to
keep attractive features of conventional GUTs (e.g.\ prediction of
hypercharge) while avoiding typical problems (e.g.\ doublet-triplet
splitting, proton decay, unrealistic fermion mass matrices). 
However, one problem of the orbifold GUTs is that they allow for lots
of freedom and one can construct many models. The number of extra
dimensions is not fixed. The grand unified group is a matter of
choice. The localisation of charged matter is not predicted but put in
by hand. That leads to the desire to construct these
phenomenologically attractive class of models from a more fundamental
theory. In the following we will demonstrate that ${\mathbb Z}_N
\times {\mathbb Z}_M$ heterotic E$_8\times$E$_8$ models naturally
contain an intermediate orbifold GUT picture. (The same is true for
${\mathbb Z}_N$ orbifolds if $N$ is not a prime number whereas for
prime numbers one always has an E$_8\times$E$_8$ symmetry in the
bulk.) 

\section{Heterotic Orbifolds}

The heterotic E$_8 \times$E$_8$ string theory is a consistent string
theory in 9+1 dimensions \cite{Gross:1985fr}. The massless spectrum
consists of gravity, E$_8\times$E$_8$ gauge fields\footnote{For recent
  discussions of the SO(32) heterotic string see\cite{so32}.} and the
  susy 
partners of minimal supersymmetry in ten dimensions. E$_8$ is an
exceptional Lie algebra with eight Cartan generators 
$H_I$, $I= 1, \ldots , 8$ and 240 root generators $E_\alpha$.
Since we do not live in 9+1 but in 3+1 dimensions we have to
compactify six dimensions. The simplest choice would be a six
dimensional torus $T^6$. In the following we will describe $T^6$ as
the direct product of three two dimensional tori $T^2$. Each $T^2$ can
be identified with a lattice in a complex plane where opposite edges of
a cell within the lattice are identified. If we just compactify on
$T^6$ we obtain an ${\cal N} =4 $ supersymmetric field theory in four
dimensions. This is not a realistic model since e.g.\ it predicts gauge
couplings which are independent of the energy scale. It would be more
realistic to have ${\cal N}=1$ supersymmetry. (No supersymmetry agrees
of course also with experimental data, but we focus on ${\cal N}=1$ in
this talk.) In order to reduce the number of supersymmetries one
replaces $T^6$ by an orbifold of $T^6$. This idea is known for a
rather long time, in the original papers
\cite{Dixon:1985jw,Dixon:1986jc}
special attention has been given to the $T^6/{\mathbb Z}_3$ orbifold.
Here, we will focus instead on $T^6/{\mathbb Z}_2\times {\mathbb Z}_2$
which should be viewed as a representative of the class $T^6/{\mathbb
  Z}_N\times {\mathbb Z}_M$ or $T^6/{\mathbb Z}_N$ with $N$ not
prime\footnote{In the free fermionic formulation, ${\mathbb Z}_2\times
  {\mathbb Z}_2$ orbifolds have been studied extensively by Faraggi and
  various collaborators, see e.g.\cite{Faraggi}.}. 
${\mathbb Z}_2\times {\mathbb Z}_2$ is the orbifold group acting on
$T^6$. Points which are mapped onto each other by the orbifold group
are identified. Our ${\mathbb Z}_2\times {\mathbb Z}_2$ orbifold group
is generated by two elements. The first element acts as a simultaneous
$180^\circ$ rotation on the first ($x^4 + i x^5$) and the second ($x^6
+ i x^7$ ) complex plane and leaves the third plane ($x^8 + i x^9$)
invariant. The second generator leaves the first plane invariant and
acts as a $180^\circ$ rotation on the second and third plane. There is
one more non trivial element in ${\mathbb Z}_2\times {\mathbb Z}_2$
which acts as a $180^\circ$ rotation on the first and third plane and
leaves the second one invariant. This element is the product of the
two generators. It is useful to associate the elements of the orbifold
group with three dimensional vectors. The two generators correspond
to
\begin{equation}
v_1 = \left( \frac{1}{2}, -\frac{1}{2}, 0\right)\,\,\, ,\,\,\,
v_2 = \left(0, \frac{1}{2}, -\frac{1}{2}\right) .
\end{equation}
This means for example that the first generator consists of a rotation
by $\frac{1}{2} \cdot 2\pi$ in the first plane, $-\frac{1}{2}\cdot
2\pi$ in the second plane and $0\cdot 2\pi$ in the third plane.

\begin{figure}
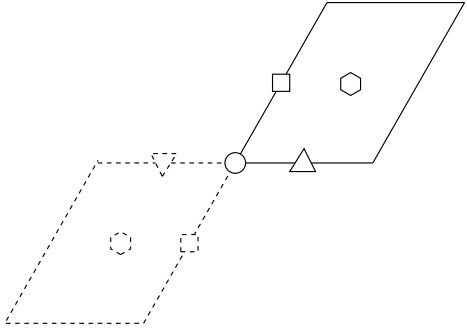

\begin{center}
\input fixed.pstex_t
\end{center}
\caption{Fixed points on one $T^2/{\mathbb Z}_2$. The $T^2$ is
  obtained by identifying opposite solid lines. Dashed lines are
  ${\mathbb Z}_2$ images of solid lines.}
\label{fig:fixed}
\end{figure}

For reasons to become clear in a moment the notion of a fixed point is
a very important concept. Let us first look at only one $T^2$ in
figure \ref{fig:fixed}. The point at the origin (depicted as a circle) 
is invariant under the rotation and trivially fixed. All other fixed
points (depicted as triangle, square and hexagon) differ from their
images by a lattice shift and thus are identical on $T^2$. The
qualitative difference between the origin and the other fixed points
will be important later. There are twisted strings which are closed
only when the orbifold identification is taken into account. The center
of mass position of these strings is constrained to the location of
a fixed point and its excitations corresponds to matter located at the
fixed point. In the ${\mathbb Z}_2 \times {\mathbb Z}_2$ orbifold each
non trivial element leaves 16 tori fixed. (Recall that on one of the
three tori the action is always trivial, the 16 counts the possible
combinations of the two sets of four fixed points in the other two
tori.) Since we have three non-trivial elements in the orbifold group
this yields 48 fixed tori altogether. 

The orbifold group acts not only on $T^6$ but will be also embedded
into the gauge group. We use the so called shift embedding
\begin{equation} \label{shiftemb}
H_I \to H_I \,\,\, ,\,\,\, E_{\vec{\alpha}} \to e^{2\pi i \vec{V}_i
  \cdot \vec{\alpha}}E_{\vec{\alpha}} .
\end{equation}
The 16 Cartan generators $H_I$ of E$_8\times$E$_8$ are left
invariant. A 16 dimensional root vector of E$_8\times$E$_8$ is denoted
by $\vec{\alpha}$ and $\vec{V}_i$ is a 16 component vector characterising
how the $i$th element of the orbifold group is embedded into the gauge
group. String theory gives very restrictive consistency conditions on
how the orbifold group can be embedded into the gauge group and it
turns out that there are only five consistent choices for the
${\mathbb Z}_2 \times {\mathbb Z}_2$ case
\cite{Forste:2004ie}. 

In the rest of the talk we will restrict ourselves to the standard
embedding which means that we copy $v_i$ into the first three
components of $\vec{V}_i$ and set the remaining components to zero
\begin{equation}
\vec{V}_i = \left( v_i,\underbrace{0,\ldots ,0}_{13}\right) .
\end{equation} 

When constructing the invariant spectrum one finds at the massless
level the gravity multiplet, the gauge multiplet and chiral
multiplets in the untwisted sector, and more chiral multiplets in the
twisted sector. In our example one finds among others (for details see
\cite{Font:1988mk, Forste:2004ie}) an E$_6\times$U(1)$^2\times$E$_8$ gauge
multiplet and one 27 dimensional representation of E$_6$ per fixed
torus. This would lead to a model with 48 generations which is to
much. In order to reduce the number of generations we need to lift the
fixed point degeneracy. 

The fixed point degeneracy is lifted when non trivial Wilson lines are
present \cite{Ibanez:1986tp}. A Wilson line is a non zero vacuum
expectation value for an internal gauge field component $A_k,
k=4,\ldots , 9$. We choose the expectation value to lie in the Cartan
subalgebra of E$_8 \times$E$_8$. This is in general not invariant under the
orbifold group (the embedding into the Cartan subalgebra is trivial but
the action on the internal dimensions is non trivial). Therefore, the
vacuum expectation value can only take discrete values such that the
orbifold image and the vev differ by a (periodic) gauge
transformation. The Wilson line is quantised. Let us look for example
at ($y$ is the fifth coordinate and $R_5$ the radius of the circle on
which it is compactified)
\begin{equation}
A_5 = \frac{i}{R_5} \vec{a}_5 \cdot \vec{H} = e^{-i\vec{a}_5 \cdot
  \vec{H} y/R_5} \partial_y  e^{i\vec{a}_5 \cdot
  \vec{H} y/R_5},
\end{equation}
which is of the form $g^{-1}\partial_y g$ and looks like a pure
gauge. This gauge is however not single valued under $y \to y + 2\pi
R_5$. We translate the vacuum expectation value for $A_5$ into the
rule that going once around the fifth direction induces the gauge
transformation
\begin{equation}
E_\alpha \to e^{2 \pi i\vec{a}_5 \cdot
  \vec{H} } E_\alpha e^{-2\pi i\vec{a}_5 \cdot \vec{H} } = e^{2\pi i
  \vec{a}_5 \cdot \vec{\alpha}} E_\alpha .
\end{equation}
Note that this looks like the shift embedding (\ref{shiftemb}). In our
case, the quantisation condition on the Wilson line is that twice $a_5$
should lie in the E$_8\times$E$_8$ root lattice. 
Now the fixed point degeneracy is lifted since the fixed points are
fixed under the 
combined action of an orbifold group element and a fixed point
dependent $T^6$ lattice shift. We should also emphasise that string
theory puts a set of consistency conditions on the Wilson lines
\cite{Ibanez:1987pj}.  

This way we can easily find a toy model with three generations
\cite{Forste:2004ie}. We use standard embedding and six non zero
Wilson lines, e.g.\
\begin{equation}
a_4 = \left(\underbrace{0,\ldots, 0}_7 ,1,1,\underbrace{0,\ldots
  ,0}_7\right) .
\end{equation}
This Wilson line breaks the E$_6$ to SO(10)$\times$U(1) (from now on
we do not 
discuss the second (hidden sector) E$_8$). The
other five Wilson lines have non zero entries only in the hidden
sector E$_8$ \cite{Forste:2004ie}.  So, in the observable sector the
gauge group is SO(10)$\times$U(1)$^3$. Because we have lifted the
degeneracy of the fixed points with the help of the Wilson lines we
now find three generations, i.e.\ three sixteen dimensional
representations of SO(10) from the twisted sectors.

\begin{figure}
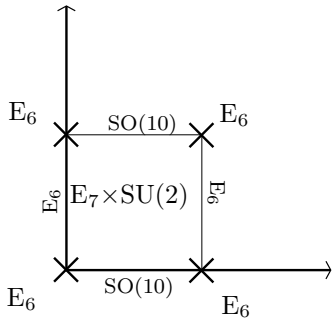

\begin{center}
\input gaugegroup_su5.pstex_t
\end{center}
\caption{Orbifold GUT with underlying heterotic orbifold.}
\label{fig:gaugegeo}
\end{figure}

Now we show that our heterotic orbifold gives rise to the picture of a
field theory orbifold given in the beginning of this talk. First we
compactify the heterotic string on a $T^4/{\mathbb Z}_2$ where the
${\mathbb Z}_2$ is generated by one of the non trivial elements of
${\mathbb Z}_2 \times {\mathbb Z}_2$ which we choose to be the second 
generator. This gives rise to a six dimensional model: four
dimensional Minkowski 
space times $T^2$. The field theory orbifold is now
obtained by replacing  $T^2$ with $T^2/{\mathbb Z}_2$, where the
${\mathbb Z}_2$ is generated by the first generator of our ${\mathbb
  Z}_2 \times {\mathbb Z}_2$. Now the gauge group in the bulk and the
embedding of the orbifold group into the field space is not a matter
of choice but given by the string theory construction. In the bulk
there is an E$_7 \times$SU(2) ($\times$ hidden sector) gauge group. At
each fixed point of $T^2/{\mathbb Z}_2$ the orbifold group imposes
projection conditions which break it to E$_6 \times$U(1)$^2$. This
looks as if the fixed points were degenerate. But for fixed points
separated in the fourth direction the E$_6$ is embedded differently
into the bulk gauge group such that the overlap is SO(10) which is the
gauge group in four dimensions, see figure \ref{fig:gaugegeo}.

\section{Conclusions}

In this talk we have demonstrated that a certain class of heterotic
string orbifolds naturally incorporate lower dimensional orbifold GUT
models. In fact, one does not need to take the detour via orbifold
GUTs but can formulate rules for string phenomenology straightaway
\cite{Nilles:2004ej}. It looks promising to find phenomenologically
interesting models by investigating the orbifold groups ${\mathbb Z}_N
\times {\mathbb Z}_M$, or ${\mathbb Z}_K$ with $K$ non prime. Wilson
lines are needed in order to obtain a realistic number of
generations. Even though heterotic orbifold constructions have been
known for quite some time the above class has found attention only
recently \cite{Kobayashi:2004ud}, \cite{Forste:2004ie},
  \cite{Kobayashi:2004ya}.  
\bigskip
                                                                              
\noindent {\bf Acknowledgments}

\noindent We would like to thank Hans
Peter Nilles and Patrick Vaudrevange for a pleasant collaboration
leading to the results presented in this talk. 
We would like to thank the organisers of String Phenomenology
2004 and the organisers of the 37th International
  Symposium Ahrenshoop on the Theory of Elementary Particles 
for creating a pleasant and stimulating atmosphere.  This work was
partially supported by the European community's Marie Curie
programs \mbox{MRTN-CT-2004-503369} ``Quest for Unification'' and
MRTN-CT-2004-005104 ``Forces Universe''.

\end{document}

%% file: orbicircle.pstex_t
\begin{picture}(0,0)%
\includegraphics{orbicircle.pstex}%
\end{picture}%
\setlength{\unitlength}{710sp}%
\begingroup\makeatletter\ifx\SetFigFont\undefined%
\gdef\SetFigFont#1#2#3#4#5{%
  \reset@font\fontsize{#1}{#2pt}%
  \fontfamily{#3}\fontseries{#4}\fontshape{#5}%
  \selectfont}%
\fi\endgroup%
\begin{picture}(10200,9942)(376,-8911)
\put(5326,-1411){\makebox(0,0)[lb]{\smash{{\SetFigFont{12}{14.4}{\rmdefault}{\mddefault}{\updefault}${\mathbb Z}_2$}}}}
\put(6301,-3361){\makebox(0,0)[lb]{\smash{{\SetFigFont{12}{14.4}{\rmdefault}{\mddefault}{\updefault}${\mathbb Z}_2^\prime$}}}}
\put(10576,-3961){\makebox(0,0)[lb]{\smash{{\SetFigFont{8}{9.6}{\rmdefault}{\mddefault}{\updefault}$\frac{\pi R}{2}$}}}}
\put(6001,539){\makebox(0,0)[lb]{\smash{{\SetFigFont{8}{9.6}{\rmdefault}{\mddefault}{\updefault}$\pi R$}}}}
\put(376,-3961){\makebox(0,0)[lb]{\smash{{\SetFigFont{8}{9.6}{\rmdefault}{\mddefault}{\updefault}$\frac{3\pi R}{2}$}}}}
\put(6001,-8911){\makebox(0,0)[lb]{\smash{{\SetFigFont{8}{9.6}{\rmdefault}{\mddefault}{\updefault}0}}}}
\end{picture}%

%% file: fixed.pstex_t
\begin{picture}(0,0)%
\includegraphics{fixed.pstex}%
\end{picture}%
\setlength{\unitlength}{947sp}%
\begingroup\makeatletter\ifx\SetFigFont\undefined%
\gdef\SetFigFont#1#2#3#4#5{%
  \reset@font\fontsize{#1}{#2pt}%
  \fontfamily{#3}\fontseries{#4}\fontshape{#5}%
  \selectfont}%
\fi\endgroup%
\begin{picture}(12066,8466)(-32,-8794)
\end{picture}%

%% file: gaugegroup_su5.pstex_t
\begin{picture}(0,0)%
\includegraphics{gaugegroup_su5.pstex}%
\end{picture}%
\setlength{\unitlength}{994sp}%
\begingroup\makeatletter\ifx\SetFigFont\undefined%
\gdef\SetFigFont#1#2#3#4#5{%
  \reset@font\fontsize{#1}{#2pt}%
  \fontfamily{#3}\fontseries{#4}\fontshape{#5}%
  \selectfont}%
\fi\endgroup%
\begin{picture}(8144,7914)(1846,-8696)
\put(7156,-3751){\makebox(0,0)[lb]{\smash{{\SetFigFont{10}{12.0}{\rmdefault}{\mddefault}{\updefault}E$_6$}}}}
\put(7201,-8521){\makebox(0,0)[lb]{\smash{{\SetFigFont{10}{12.0}{\rmdefault}{\mddefault}{\updefault}E$_6$}}}}
\put(1846,-8341){\makebox(0,0)[lb]{\smash{{\SetFigFont{10}{12.0}{\rmdefault}{\mddefault}{\updefault}E$_6$}}}}
\put(1891,-3706){\makebox(0,0)[lb]{\smash{{\SetFigFont{10}{12.0}{\rmdefault}{\mddefault}{\updefault}E$_6$}}}}
\put(3421,-5776){\makebox(0,0)[lb]{\smash{{\SetFigFont{10}{12.0}{\rmdefault}{\mddefault}{\updefault}E$_7\times$SU(2)}}}}
\put(3106,-6046){\rotatebox{90.0}{\makebox(0,0)[lb]{\smash{{\SetFigFont{8}{9.6}{\rmdefault}{\mddefault}{\updefault}E$_6$}}}}}
\put(4366,-3931){\makebox(0,0)[lb]{\smash{{\SetFigFont{8}{9.6}{\rmdefault}{\mddefault}{\updefault}SO(10)}}}}
\put(6931,-5191){\rotatebox{270.0}{\makebox(0,0)[lb]{\smash{{\SetFigFont{8}{9.6}{\rmdefault}{\mddefault}{\updefault}E$_6$}}}}}
\put(4276,-8026){\makebox(0,0)[lb]{\smash{{\SetFigFont{8}{9.6}{\rmdefault}{\mddefault}{\updefault}SO(10)}}}}
\end{picture}%